\documentstyle[epsf]{l-aa}
\newcommand{\subs}[1]{\mbox{\scriptsize #1}}
\newcommand{\stru}[1]{\rule {0ex}{#1}}
\begin{document}
\thesaurus{2(11.17.4 HE\,2347$-$4342; 11.17.1; 12.03.3)}
\title
{Patchy intergalactic He\,{\sc ii} absorption in HE\,2347$-$4342
\thanks{Based on observations collected at the European Southern Observatory, La
Silla, Chile (ESO No.\ 58.B--0116). Based on IUE observations collected at the ESA VILSPA ground station near Madrid, Spain.
Based on observations with the NASA/ESA Hubble Space Telescope, obtained at
the Space Telescope Science Institute, which is operated by Aura, Inc., under
NASA contract NAS 5--26\,555. }}
\subtitle{The possible discovery of the epoch of He-reionization}
\author {D. Reimers \inst{1}
\and S. K\"ohler \inst{1}
\and L. Wisotzki \inst{1}
\and D. Groote \inst{1}
\and P. Rodriguez-Pascual \inst{2}
\and W. Wamsteker \inst{2}}
\institute{Hamburger Sternwarte, Universit\"at Hamburg, Gojenbergsweg 112,
D-21\,029 Hamburg \and
ESA-Vilspa, PO Box 50727, E-28\,080 Madrid}
\offprints{D.~Reimers, \protect \\dreimers@hs.uni-hamburg.de}
\date{received date; accepted date}
\maketitle
\markboth{D.~Reimers et al.: He\,{\sc ii} absorption in HE\,2347$-$4342}{}
\begin{abstract}
We report on observations of redshifted He\,{\sc ii}\,303.8~\AA\ 
absorption in the  high-redshift QSO
HE\,2347$-$4342 ($z=2.885$, $V=16.1$) with the Goddard High
Resolution Spectrograph on board HST in its low resolution
mode ($\bigtriangleup\lambda= 0.7$ \AA).
With $f_{\lambda}=3.6\ 10^{-15}$ erg\,cm$^{-2}$\,s$^{-1}$\,\AA$^{-1}$ at the 
expected position of He\,{\sc ii}\,304 \AA\ absorption it is the most UV-bright high redshift QSO discovered so far.

We show that the He\,{\sc ii} opacity as a function of redshift is
patchy showing spectral regions with low He\,{\sc ii} opacity (``voids'') 
and regions with
high He\,{\sc ii} opacity (blacked-out ``troughs'') and 
no detectable flux. Combination with high-resolution optical spectra of 
the Ly$\alpha$ forest using CASPEC at the 3.6\,m telescope shows that  
the voids can be explained either exclusively by Ly$\alpha$ forest 
cloud absorption with a moderate 
$N_{\subs{He\,{\sc ii}}}/N_{\subs{H\,{\sc i}}}$ ratio $\eta\leq\,100$ and
turbulent line broadening or by a combination of Ly$\alpha$ forest
with $\eta$ = 45 and thermal broadening plus a diffuse medium with
$\tau_{\subs{GP}}^{\subs{He\,{\sc ii}}} \approx$ 0.3. Since the latter is 
a minimum assumption for the Ly$\alpha$ forest,  a strict upper limit to a
 diffuse medium is $\Omega_{\subs{diff}}<0.02$ h$_{50}^{-1.5}$ at $z=2.8$.

In the troughs in addition to the Ly$\alpha$
forest opacity a continuous He\,{\sc ii} \mbox{304 \AA} opacity 
$\tau = 4.8^{+\infty}_{-2}$
is required. In case of photoionization, the troughs would
require a diffuse component with a density close to $\Omega
\simeq\,0.077\,(\eta/45)^{-0.5}$ h$_{50}^{-1.5}$, i.e.\  all baryons in the universe, which is inconsistent,
however, with the observed absence of such a component in the voids.
A tentative interpretation is that we observe the epoch of partial
He\,{\sc ii} reionization of the universe with patches not yet reionized.
In that case a diffuse component with 
$\Omega_{\subs{diff}}\geq\, 1.3\ 10^{-4}$ 
h$^{-1}_{50}$
would be sufficient to explain the ``trough'' opacity.
The size of the 1163--1172 \AA\ trough is  $\sim\,6$ h$_{50}^{-1}$ Mpc
 or $\sim\,2300$ km\,s$^{-1}$, respectively.

We also discuss partially resolved He\,{\sc ii} absorption of a high-ionization
associated absorption system. Despite its high luminosity HE\,2347$-$4342 does not show a He\,{\sc ii} proximity effect. A possible reason is that the strong
associated system shields the He\,{\sc ii} ionizing continuum.
\end{abstract}

\section{Introduction}
A longstanding goal of observational cosmology has been the detection of
a diffuse intergalactic medium (IGM), suspected to contain a major
fraction of the baryons of the universe produced in the big-bang, 
by means of the Gunn-Peterson (1965) effect. The test using H\,{\sc i} Ly$\alpha$
has been negative so far (e.g.\ Steidel \& Sargent 1987; Giallongo 
et al.\ 1992, 1994). The conclusion has been that the IGM is highly
ionized and/or contains a much lower fraction of the baryons of the
universe than originally expected. Recent hydrodynamical models of
structure formation indeed predict that the fragmentation of baryons is
nearly complete and that a diffuse IGM is suppressed by two
orders of magnitude (Miralda-Escud\'{e} et al.\ 1996; Meiksin 1997).

With the launch of the Hubble Space Telescope it was hoped to detect
a highly ionized diffuse IGM via the He\,{\sc ii} Ly$\alpha$ (303.8 \AA)
line, which was predicted to be much more sensitive
than H\,{\sc i} Ly$\alpha$.

The observation of absorption on the blue side of the He\,{\sc ii} 304 \AA\ line
in high-redshift QSOs has been pioneered by Jakobsen et al.\ (1994) who
observed a completely absorbed spectrum ($\tau =3.2^{+\infty}_{-1.1}$) on the blue side of the
redshifted He\,{\sc ii} line in Q\,0302$-$003 ($z=3.29$). 
He\,{\sc ii} absorption in a second QSO, 
 PKS\,1935$-$692
($z=3.18$), was discovered with HST by Tytler (1995), with a similar result
(cf.\ Jakobsen 1996), while Davidsen et al.\ (1996) observed a lower He\,{\sc ii}
opacity ($\tau = 1.00\pm\,0.07$ at $\bar{z} = 2.4$) toward HS\,1700+6416 ($z=2.72$) with
HUT/ASTRO-2.

Due to their low spectral resolution, all these observations did not allow to
distinguish between a He\,{\sc ii}\,304 \AA\ forest and absorption by a diffuse
medium. By means of model calculations using high resolution optical
H\,{\sc i} Ly$\alpha$ forest spectra, both Songaila et al.\ (1995) for Q\,0302$-$003
and Davidsen et al.\ (1996) for HS\,1700+6416 demonstrated that the He\,{\sc ii}
opacity could be explained by the Ly$\alpha$ forest alone with
$N_{\subs{He\,{\sc ii}}}/N_{\subs{H\,{\sc i}}}\simeq$ 80, a value roughly consistent with photoionization
calculations using predicted metagalactic radiation fields due to QSOs
with absorption by intergalactic matter taken into account (Meiksin \& Madau 1993;
Giroux et al.\ 1995; Haardt \& Madau 1996). On the other hand, recent
reobservations of Q\,0302$-$003  with the GHRS on board of HST by Hogan
et al.\ (1997) seem to indicate that a diffuse component with
 $\Omega_{\subs{diff}} \simeq$\,0.01 (h/0.7)$^{-1.5}$
is required in order to explain the He\,{\sc ii} opacity in the 
well-known
``void'' (cf. Dobrzycki \& Bechthold 1991) in the \mbox{Ly$\alpha$} forest 
of Q\,0302$-$003.
All the HST observations of the intergalactic He\,{\sc ii} opacity obtained so far suffered
from the faintness ($V>18$) of the QSOs. The difficulty in finding more
suitable targets is best demonstrated by the fact that the FOC surveys
for unabsorbed $z>3$ QSOs by Jakobsen et al.\  and Tytler et al.\ detected only
2 moderate opacity lines of sight in more than 110 observed QSOs
(cf.\ Jakobsen 1996).
In this paper we report on the discovery of the extremely bright ($V=16.1$),
 $z=2.885$ QSO HE\,2347$-$4342 within the Hamburg/ESO survey for
bright QSOs
and on successful observations of He\,{\sc ii}\,304~\AA\ absorption with the GHRS
onboard the Hubble Space Telescope in its low re\-solution mode. The
combination of partially resolved He\,{\sc ii} absorption with high resolution 
optical spectra taken with CASPEC at the 
ESO 3.6\,m telescope allows to constrain the 
$N_{\subs{He\,{\sc ii}}}/N_{\subs{H\,{\sc i}}}$ ratio in \mbox{Ly$\alpha$}
forest clouds and to quantify the contribution of a diffuse component to the 
He\,{\sc ii} absorption.

\section{Observations}

\subsection{Discovery and IUE observations}
\begin{figure*}[t]
\epsfclipon
\epsffile[40 280 544 604]{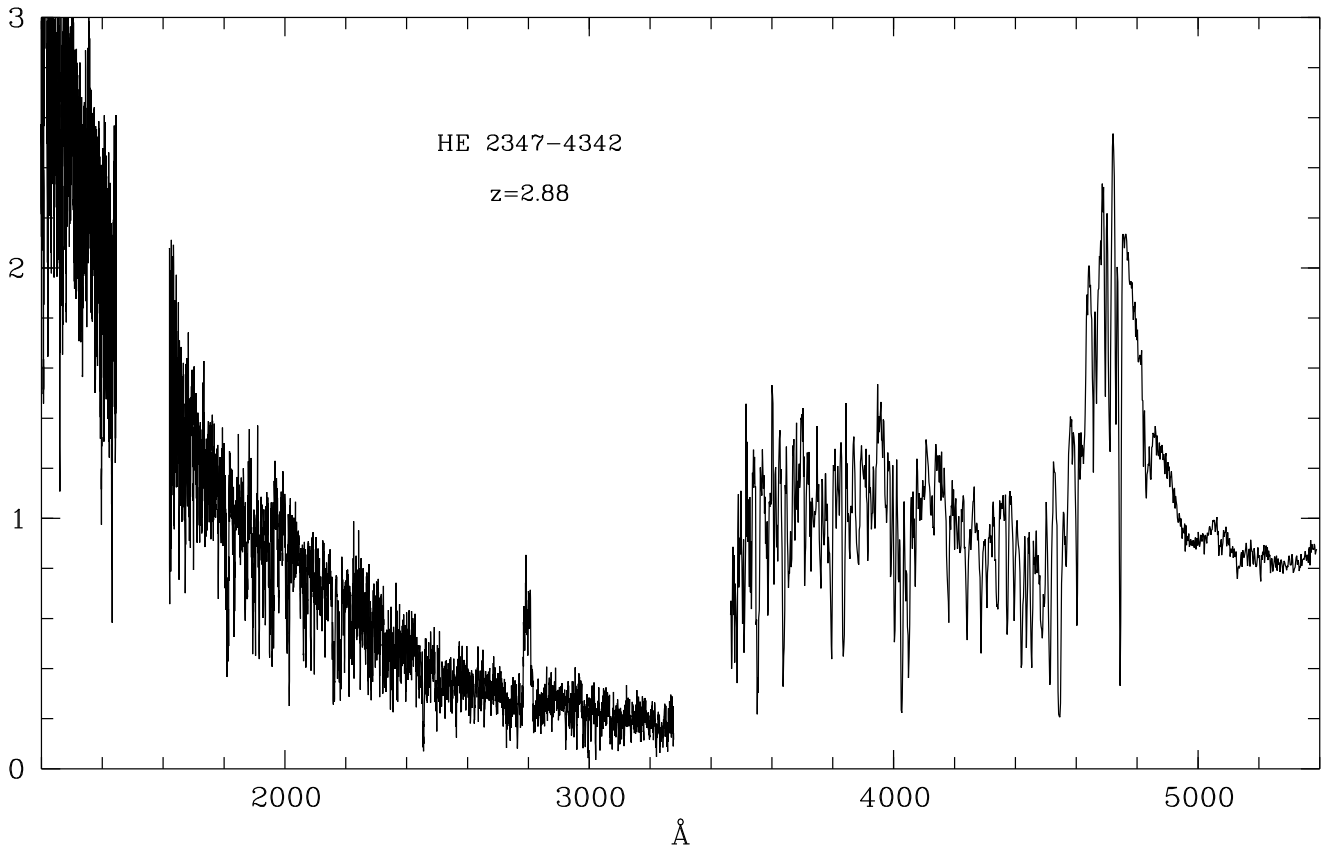}
\caption{Combined low-resolution optical spectrum observed with the ESO 2.2\,m telescope and ultraviolet spectra 
obtained with the FOS and GHRS onboard the HST. Flux is given in units of 
10$^{-15}$ erg s$^{-1}$ cm$^{-2}$ \AA$^{-1}$. The feature at 2800 \AA\ is an artefact.}
\label{fhstopt}
\end{figure*}

The bright high redshift QSO HE\,2347$-$4342 was discovered           
as part of the Hamburg/ESO Survey for bright QSOs a wide-angle survey based on objective-prism plates taken with
the ESO 1\,m Schmidt telescope. The plates are digitized in Hamburg and
automatically searched for QSO candidates which are subsequently observed 
spectroscopically  with ESO telescopes (Reimers 1990; Wisotzki et al.\ 1996;
Reimers et al.\ 1996). HE\,2347$-$4342 was confirmed as an extremely bright
($V=16.1$), high-redshift ($z=2.88$) QSO in an observing run at the ESO
2.2\,m telescope in October 1995.
The coordinates are 23$^{h}$50$^{m}$34.3$^{s}$ $-43\degr\,26'\,00''$ (2000).
The optical spectrum, spectral resolution $\sim\,8$ \AA, is shown in 
Fig.~\ref{fhstopt}. Because of the absence of damped Ly$\alpha$ lines and Lyman limit systems, HE\,2347$-$4342 was immediately recognized as a
candidate suitable for UV follow-up observations.

Another  low resolution spectrum (20 \AA) was taken 
in October 
1996 with the ESO 1.5\,m telescope covering the wavelength range $\lambda\lambda$
3200--9000 \AA\ with the aim to obtain an improved QSO redshift.
We find $z=2.870\pm$ 0.005 from the C\,{\sc iii} and C\,{\sc iv} lines.
 However,  we find a higher redshift of
$z=2.885\pm$0.005 in both  low- and high-resolution spectra of the O\,{\sc i} line.
Since it is well known that the higher ionization lines underestimate the intrinsic QSO redshift, we shall use 2.885 hereafter.

HE\,2347$-$4342 was observed twice with the Short Wavelength Prime (SWP) camera 
onboard IUE (SWP 56218, 580$^{m}$, 56228, 712$^{m}$).
The QSO was detected in both images, although with very low 
signal-to-noise ratio  at the 1--2\ 10$^{-15}$ erg cm$^{-2}$ s$^{-1}$ 
\AA$^{-1}$ level. 

With a flux more than a factor of 10 higher at the expected wavelength
of the redshifted He\,{\sc ii} 304 \AA\ line than in Q\,0302$-$003 and
PKS\,1935$-$692  (cf.\ Jakobsen 1996), HE\,2347$-$4342 offered the chance to observe the He\,{\sc ii}
Ly$\alpha$ forest with the GHRS and the hope of resolving 
 the He\,{\sc ii} forest.

\subsection{UV observations with HST}
The ultraviolet spectra have been taken in three visits between June 7 
and June 14, 1996 with both the FOS in its high-resolution mode (R=1300)
and the GHRS in its low resolution mode.
The log of observations is given in Table\,1.
The standard pipeline processing provided flux calibrated data together with the 1$\sigma$ error in
the flux of each pixel as a function of wavelength. The maximum
signal-to-noise ratio
achieved for the GHRS observation is 14. With the GHRS first-order grating the
background  is mainly due to counts
from particle radiation. Since the background is usually very low the average
over all diodes is subtracted from the science data. Inspecting the raw science
and background data we find a mean background level of 1--2 \% of the mean quasar countrate.

\begin{table}\caption{\bf The log of HST observations}
\begin{tabular}{ccccc}

Detector/ & Spectral & Aperture & Resolution & Exposure\\ 
Grating &   range    &          & FWHM & time (s)\\
\hline
Red\,G270H & 2220--3240 & 4$\farcs$3 & 2 \AA  & 1320\stru{3ex} \\

Red\,G190H & 1620--2240 & 4$\farcs$3 & 1.5 \AA  & 7930 \\

    G140L & 1150--1440 & LSA & 0.7 \AA  & 21106 \\  
\end{tabular}
\end{table}

The overall spectral energy distribution is shown in Fig.~\ref{fhstopt}, together with a 
low resolution optical spectrum.

The strong break in the spectrum near 3500 \AA\ is due to a Lyman limit system at $z=2.739$ 
with an optical depth \mbox{$\tau = 1.6,$} from which the flux recovers 
($\tau_{\nu} \propto\,\nu^{-3}$) and rises to a maximum continuum flux at $\sim\,1180$\,\AA\ 
of $\sim\, 3.6\,\cdot\, 10^{-15}$ erg\,s$^{-1}$\,cm$^{-2}$\,\AA$^{-1}$
just longward of the 
He\,{\sc ii} break. HE\,2347$-$4342 is nearly a factor of 2 brighter 
than HS\,1700+6416 
(Reimers et al.\ 1992) at the shortest wavelengths making it the most UV-bright 
high redshift quasar discovered so far.

In this paper we discuss only the strong He\,{\sc ii} absorption shortward of 1186 
\AA. A detailed analysis of the rich QSO absorption line spectrum will be 
deferred to a future paper in which also the high-resolution optical spectra 
will be included.

\begin{figure*}[t]
\epsfclipon
\epsffile[40 280 544 604]{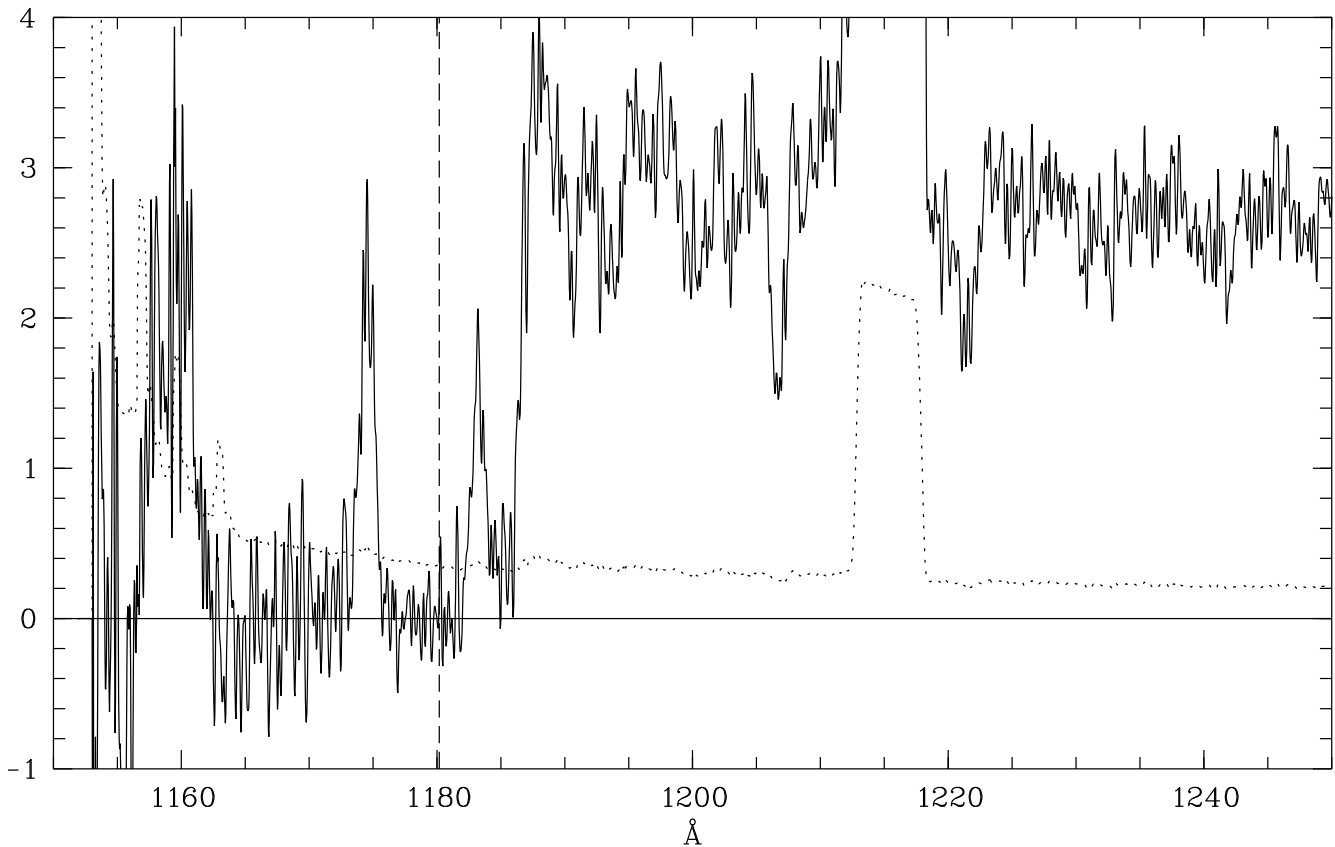}
\caption{Section of the GHRS spectrum and error spectrum  (dotted curve) 
of HE\,2347$-$4342 with the flux given in 10$^{-15}$ 
erg\,s$^{-1}$\,cm$^{-2}$\,\AA$^{-1}$. The expected position of the 
He\,{\sc ii}\,303.78 \AA\ edge for a QSO redshift 
of $z=2.885$ is indicated by the vertical dashed line. }
\label{fheiiedge}
\end{figure*}

Figure~\ref{fheiiedge} shows the relevant part of the GHRS spectrum. The absolute wavelength 
scale was checked using the strong interstellar  lines  Si\,{\sc ii}\,1260 and
C\,{\sc ii}\,1335. 
We applied a wavelength zero point offset of $-$0.15 \AA\  to shift the
interstellar absorption lines to  their rest wavelengths.

The first question we ask is whether the observed strong He\,{\sc ii}\,303.78\,\AA\ edge 
is at the expected position. With an observed QSO redshift 
 from the O\,{\sc i}\,1302 emission
line in
both low and high resolution spectra of $z=2.885\pm\,0.005$, we expect
the He\,{\sc ii} edge at 1180.2 $\pm$ 1.5 \AA. The observed break is $\sim\,6$
\AA\ to the red of this.

\begin{figure}[t]
\epsfclipon
\epsffile[46 377 284 738]{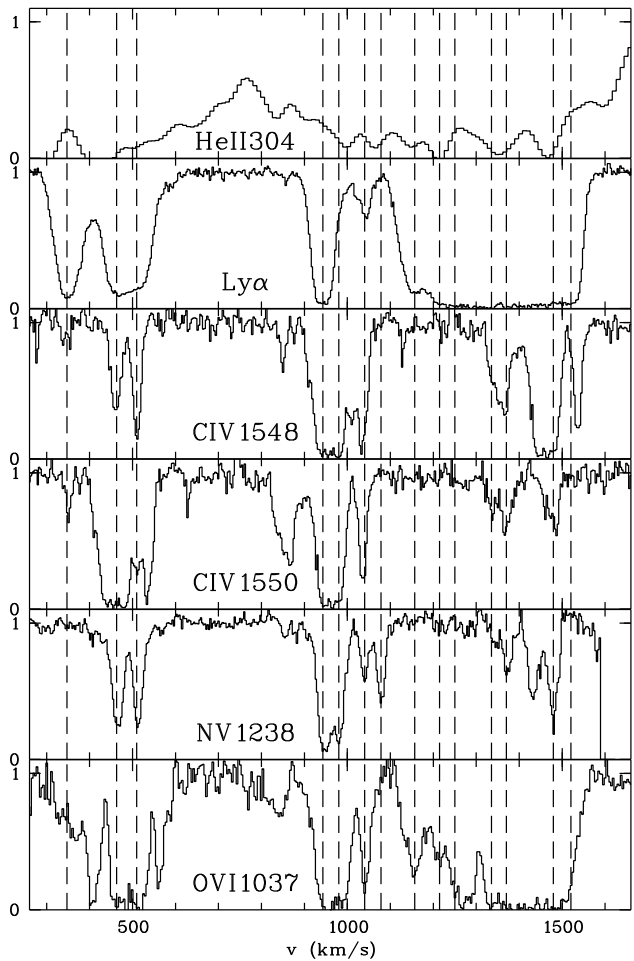}
\caption{Velocity profiles of absorption lines arising in the associated
system as seen in the normalized high-resolution optical data. The velocity
$v=0$ km\,s$^{-1}$ corresponds to a QSO redshift $z=2.885$. At
least 14 individual components can be identified by heavy element and/or
H\,{\sc i} Lyman series absorption lines. The complex associated system is
responsible for the observed He\,{\sc ii}\,303 \AA\ absorption in HST data
(cf.\ top panel) redward of the expected 
He\,{\sc ii}\,303 \AA\ edge at 1180$\pm$1.5~\AA\ 
(corresponding to v$\,=0$ km\,s$^{-1}$). }
\label{fassozi}
\end{figure}

The explanation for this discrepancy is the presence of a strong 
multicomponent associated ($z_{\subs{abs}}\simeq\,z_{\subs{em}}$) absorption system with redshifts 2.8911, 2.8917, 
2.8972, 2.8977, 2.8985, 2.8989, 2.9023, 2.9028 and 2.9041. This associated system 
is extremely strong in H\,{\sc i} and  in  O\,{\sc vi} and strong in N\,{\sc v}, 
C\,{\sc iv} and O\,{\sc v} (cf.\ Fig.~\ref{fassozi}). In He\,{\sc ii}\,303.78 two 
absorption complexes can be seen, the broad one ranging from 1184 to 1187 \AA\ and the 
unresolved line pair $z=2.891$ at 1182 \AA\ (Fig.~\ref{fheiihi}). The former is not saturated 
in  He\,{\sc ii}, in contrast to H\,{\sc i} and  O\,{\sc vi}, which
means that He\,{\sc ii} is highly ionized 
close to the QSO. The apparent He\,{\sc ii} edge at \mbox{1186.5 \AA} and the 
absorption 
between 1182 and 1186.5~\AA\ is exclusively due to the partially resolved 
He\,{\sc ii} lines of the associated system. Besides the highly redshifted associated 
systems with $z=2.891$ to 2.904 (up to 1500 km\,s$^{-1}$ relative to the QSO at $z=2.885$) there 
are two further systems with highly ionized species: $z=2.878$ (N\,{\sc v} and
C\,{\sc iv}, unsaturated Ly$\alpha$) and 2.863 (O\,{\sc vi},
weak C\,{\sc iv}, no N\,{\sc v}, saturated 
Ly$\alpha$). The latter again appears to show an unsaturated He\,{\sc ii}\,303.8\,\AA\ 
line, like the redshifted associated system.

In summary, if the associated He\,{\sc ii} absorption is taken into account, the true 
observed He\,{\sc ii} edge is estimated to be around $z=2.889$, cf.\ Fig.~\ref{fassozi}, close to the QSO emission redshift
$z=2.885\pm\,0.005$. 

\subsubsection{Continuum definition}
The GHRS data were first corrected for interstellar reddening
according to Seaton's law (Seaton 1979) with $E(B-V)=0.0387$,
corresponding to N(H\,{\sc i})=\,2.01 10$^{20}\,$cm$^{-2}$ (Stark et al.\
1992).  A local continuum definition
is difficult due to the high absorption line density.
 We searched for regions in
the data apparently free of absorption lines, where we calculated the mean
flux and the error in the mean flux to check for consistency with the noise.
The continuum for the GHRS data was then constructed by fitting
a low-order polynomial  to these mean flux values. As is obvious from the GHRS
data longward of the He\,{\sc ii} edge there are still
numerous absorption lines   presumable from heavy elements, since the number
of Ly$\alpha$ clouds expected in this range is small at these low redshifts
(Bahcall et al.\ 1993). Thus the
derived continuum level might
be underestimated due to unresolved line blending. This effect, however, 
 should not influence the
analysis of the He\,{\sc ii} absorption since it is expected to be of the same
amount longward and shortward of the He\,{\sc ii} edge.

\subsection{Optical Observations and Data Reduction}
 HE\,2347$-$4342 was observed on two nights  (4--6 October 1996) using CASPEC
with its Long Camera  on the ESO 3.6\,m telescope at La Silla. We obtained
three individual 2h45m exposures in the spectral range from 3550 to 4830~\AA\
and a single 2h45m exposure in the range from 4870 to 6180 \AA. 
The slit was aligned with the parallactic angle in order to minimize light
losses due to atmospheric dispersion.  
The data reduction was done with the ECHELLE software package available in 
  MIDAS, supplemented by own software
programmes for an
optimal extraction of the echelle orders (kindly provided by
S.\ Lopez).

For wavelength calibration, a Th-Ar comparison spectrum was obtained
immediately before  and after each QSO observation. For each order the wavelength scale
was determined by fitting a third-order polynomial to the automatically identified lines.
The resulting rms residuals were
0.003/0.005 \AA\ for the two spectral regions.
The individual observations were rebinned to the same wavelength scale and
each observation was 
scaled by its median and then coadded,  weighting by
the inverse variance.
After correction for the  blaze function using observations of the standard
star $\mu$Columbae,
  the continuum for each order was determined by fitting low-order
polynomials to regions free of absorption lines.
Wavelengths have been corrected to vacuum, heliocentric values.
The final resolution achieved is R=21\,500 with a signal-to-noise ratio of 34 per pixel at
$\lambda_{\subs{obs}}$=4700 \AA. For the single exposure longward of 4870 \AA\ we reached R=24\,500 and a
S/N=14.

For a quantitative analysis of the He\,{\sc ii} edge we need to identify the
corresponding H\,{\sc i} Ly$\alpha$ clouds in the optical data.
All absorption lines were fitted with Voigt
profiles convolved with the instrumental profile  using the software package FITLYMAN
available in MIDAS  to derive $z$, $N_{\subs{H\,{\sc i}}}$ and $b$ values. 
 For the Ly$\alpha$ absorber clouds  responsible for the
observed part of the He\,{\sc ii} edge the optical data cover also Ly$\beta$, Ly$\gamma$
and Ly$\delta$ absorption lines allowing either an independent derivation of
$z$, $N_{\subs{H\,{\sc i}}}$
and $b$ or at least a consistency check with the fit results for Ly$\alpha$.
With the available optical data we can detect Ly$\alpha$ absorption lines by
neutral hydrogen down to column densities log\,$N_{\subs{H\,{\sc i}}}=12.6$.

\begin{figure*}[t]
\epsfclipon
\epsffile[40 280 544 604]{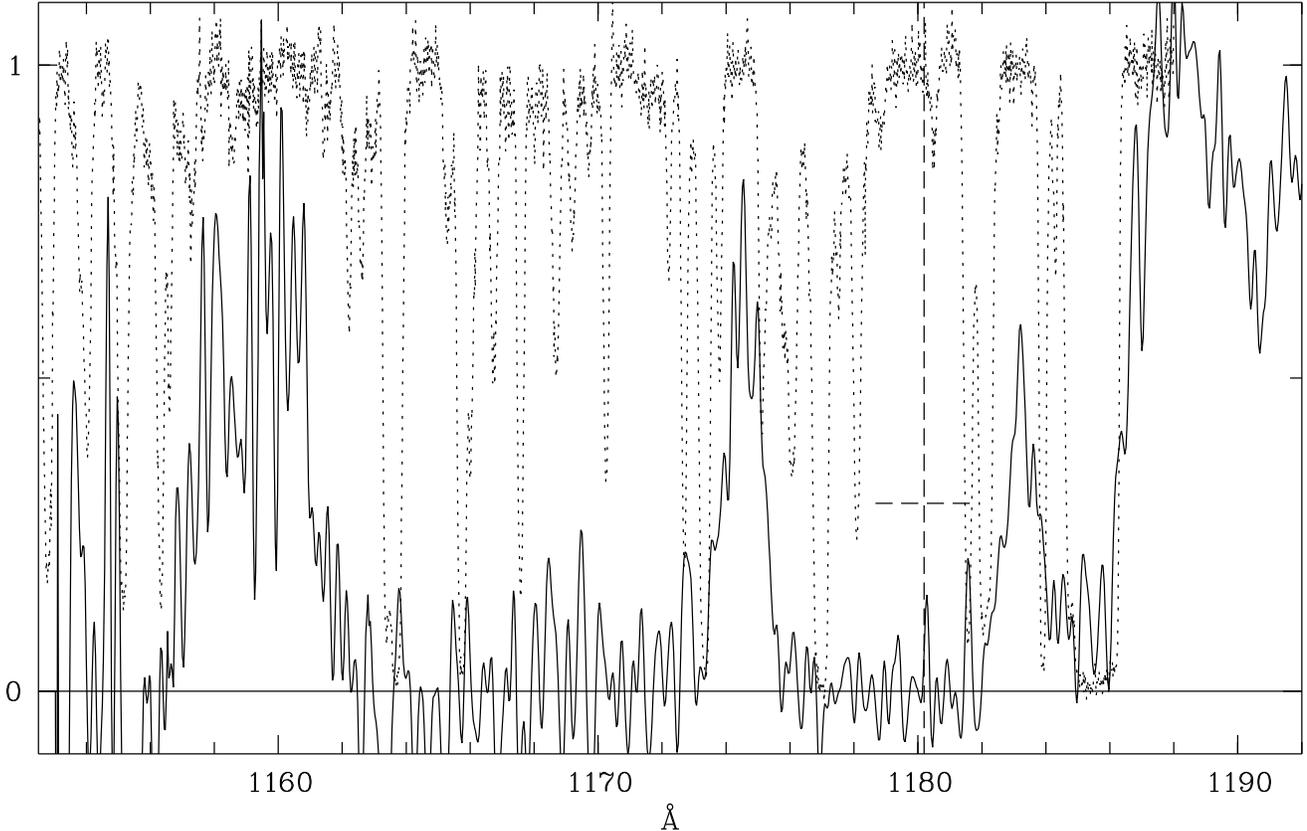}
\caption{He\,{\sc ii}\,304 \AA\ forest as seen in the normalized GHRS 
spectrum overlaid 
with the corresponding H\,{\sc i} Ly$\alpha$ forest in the normalized 
high-resolution optical data (dotted curve) scaled in wavelength according to 
303.78/1215.67.  The expected position of the 
He\,{\sc ii}\,303.78 \AA\ edge according to a QSO redshift 
of $z=2.885\pm0.005$ is indicated by the dashed lines.}
\label{fheiihi}
\end{figure*}

\section{Interpretation}

Are the observed Ly$\alpha$ forest clouds 
the main source of the observed strong He\,{\sc ii} opacity? This is 
addressed in Fig.~\ref{fheiihi} where the normalized high-resolution optical Ly$\alpha$
forest spectrum (scaled in wavelength according to 303.78/1215.67) 
is overlayed  the normalized He\,{\sc ii} ``forest'' spectrum. 

We note three different types of He\,{\sc ii} absorption regions:

\begin{itemize}
\item[--] the associated systems ($z = 2.891$ to 2.904), 
which are strong in C\,{\sc iv}, N\,{\sc v}, O\,{\sc vi}, and O\,{\sc v} lines, can clearly be 
recognized in He\,{\sc ii}. The He\,{\sc ii} lines are partially resolved and
 weaker 
than H\,{\sc i} Ly$\alpha$.

\item[--] there are two ``voids'' in the Ly$\alpha$ forest spectrum which 
are also seen in He\,{\sc ii}: a void-like structure similar to the one in the
Ly$\alpha$  forest of Q\,0302$-$003 around $4\times\,1160$~\AA\ with a width of $\sim\,20$~\AA\ 
(5~\AA\ in He\,{\sc ii}) and a further void at $\sim\,4 \times 1174.5$ \AA.

\item[--] in the remaining part of the He\,{\sc ii} spectrum, in the
\mbox{``troughs''} at 
$\lambda\lambda$ 1163--1172 \AA\ and 1176--1181 \AA, there is no
detectable flux in the He\,{\sc ii} forest ($\tau =  4.8^{+\infty}_{-2}$), and in particular
there is no relation to the H\,{\sc i} Ly$\alpha$ forest, in spite of several
smaller ``voids'' in the Ly$\alpha$ forest (at $4\,\times\,\sim\,1180$, 1171,
1164.5 \AA). The 1163--1172 \AA\ trough has a size of 
$\sim\,6$ h$_{50}^{-1}$ Mpc (comoving) 
or 2300 km\,s$^{-1}$.
\end{itemize}

We have used the column densities, $b$-values and redshifts of the detected H\,{\sc i} clouds to predict
the He\,{\sc ii} absorption, adopting either turbulent or thermal line
broadening.
Simulated spectra were degraded to the GHRS resolution for a comparison with
the observation.
It is clear already from Fig.~\ref{fheiihi} that the three components
 (``voids'', ``troughs'' and the associated
system) cannot be modelled by scaling the Ly$\alpha$ forest clouds
with one constant column density ratio $\eta =
N_{\subs{He\,{\sc ii}}}/N_{\subs{H\,{\sc i}}}$.

\begin{figure*}[t]
\epsfclipon
\epsffile[40 280 544 604]{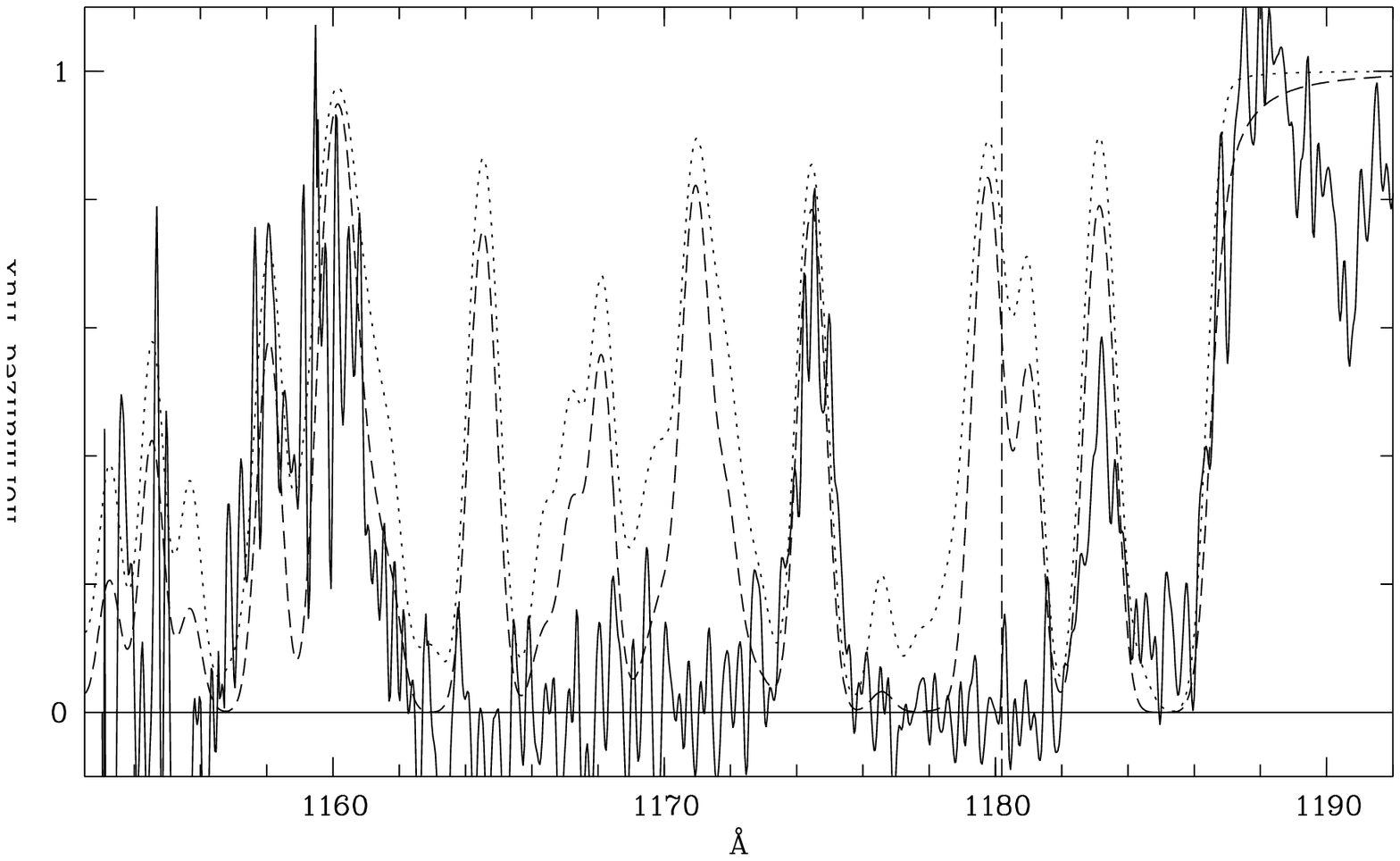}
\caption{The normalized GHRS spectrum overlaid with the  model He\,{\sc ii} absorption spectrum 
predicted on the basis of redshifts, H\,{\sc i} column densities and 
velocity dispersion parameters of the   Ly$\alpha$ forest lines detected 
in the optical data.
 The dotted (dashed) curve corresponds to the assumptions 
$N_{\subs{He\,{\sc ii}}}/N_{\subs{H\,{\sc i}}}=100$ (1000) and 
pure turbulent broadening $b_{\subs{He\,{\sc ii}}}=b_{\subs{H\,{\sc i}}}$. }
\label{fheiinorm}
\end{figure*}

This is shown in Fig.~\ref{fheiinorm} where the model calculation assumes
$\eta$ = 100 and $b_{\subs{He\,{\sc ii}}} = b_{\subs{H\,{\sc i}}}$, i.e.\ pure turbulent broadening, which
produces  maximum  He\,{\sc ii} opacity. While the two ``voids'' around 1160
and 1174 \AA\ are perfectly matched - notice the structure within
the 1160 \AA\ void and the strong line on the short wavelength side
of the 1174 void - the troughs cannot be modelled with the
observed Ly$\alpha$ forest clouds, even for
$\eta$ = 1000. A continuous H\,{\sc i} component with $\tau$ = 0.01 or a
quasicontinuous blend of weak lines like the ``undulating'' absorption
observed by Tytler (1995) in high S/N spectra of HS\,1946+7658 could
yield the observed He\,{\sc ii} opacity for $\eta\,>1000$. However, whatever
combinations of $\eta$ and broadening parameters $b$ are chosen, the ``voids''
and ``troughs'' cannot be modelled with the same parameter set.

As was  shown already by Hogan et al.\ (1997), the observed Ly$\alpha$ clouds in
the high-resolution optical spectrum of Q\,0302$-$003 alone
cannot explain the absorption even for extremely soft ionizing spectra.
Since our optical data do not allow the detection of hydrogen clouds down to
log\,$N_{\subs{H\,{\sc i}}}=12$ cm$^{-2}$, we performed Monte Carlo simulations to
estimate the amount of He\,{\sc ii} absorption by the weakest Ly$\alpha$
forest clouds. The distribution of Ly$\alpha$ clouds in redshift and column
density  was described by
\begin{equation}
\frac{\delta^{2}N}{\delta\,N_{\subs{HI}}\,\delta\,z}=
A\,(1+z)^{\gamma}\,N_{\subs{HI}}^{\beta}
\end{equation}
chosing A=2.4\,10$^{7}$, $\gamma=2.46$, $\beta=-1.5$ and $N_{\subs{min}}=2\,10^{12}$
cm$^{-2}$ (e.g.\ Madau 1995
and references therein). But even for  $\eta=1000$ and turbulent broadening
we reach the same conclusion as Hogan et al.\ (1997).

As far as the associated system is concerned, we expect a much harder
ionizing spectrum due to the quasar, i.e.\ lower values of $\eta$.
Considering both heavy element and hydrogen absorption lines in the optical
data we can distinguish 14
individual components of the associated system ranging from $z=2.8896$ to
2.9047, i.e.\ spanning $\approx\,1000$ km\,s$^{-1}$.   For absorber components with the
strongest H\,{\sc i} absorption (i.e.\ those at
$z>2.8985$) the observed He\,{\sc ii} absorption is best modelled adopting $\eta=5$ for
$b_{\subs{He\,{\sc ii}}}=b_{\subs{H\,{\sc i}}}$, while at lower redshift ($z=2.8896$--2.8985) $\eta=100$ is more appropriate. 
According to our estimates for the spectral shape of the intrinsic quasar energy
distribution (see below) we would expect  $\eta\approx\,8$.

We conclude from the above  that  
the ionization structure of the associated systems also
 cannot be modelled by a single set of parameters 
(cf.\ discussion in Sect.~3.4).

\subsection{``Voids''}

\subsubsection{Pure Ly$\alpha$ forest opacity}

Figure~\ref{fheiinorm} shows that
the ``void'' at 1160 \AA\ can be explained by
photoionized forest clouds far from the QSO with $\eta = 100$ and
turbulent broadening $b_{\subs{He\,{\sc ii}}}= b_{\subs{H\,{\sc i}}}$.
This high value of $\eta$ is barely consistent with models of an ionizing metagalactic
background due to QSOs which predict $\eta = 45$
(Haardt \& Madau 1996). However, the
faint end of the QSO luminosity function is not known for $z > 3$
and there may be a contribution by stars  (Rauch et al.\ 1997).
This ``void'' component is consistent with what Davidsen et al. (1996)
found at lower redshift $z=2.4$ in HS\,1700+6416.

\subsubsection{Ly$\alpha$ forest plus diffuse component}
\begin{figure*}[t]
\epsfclipon
\epsffile[40 280 544 604]{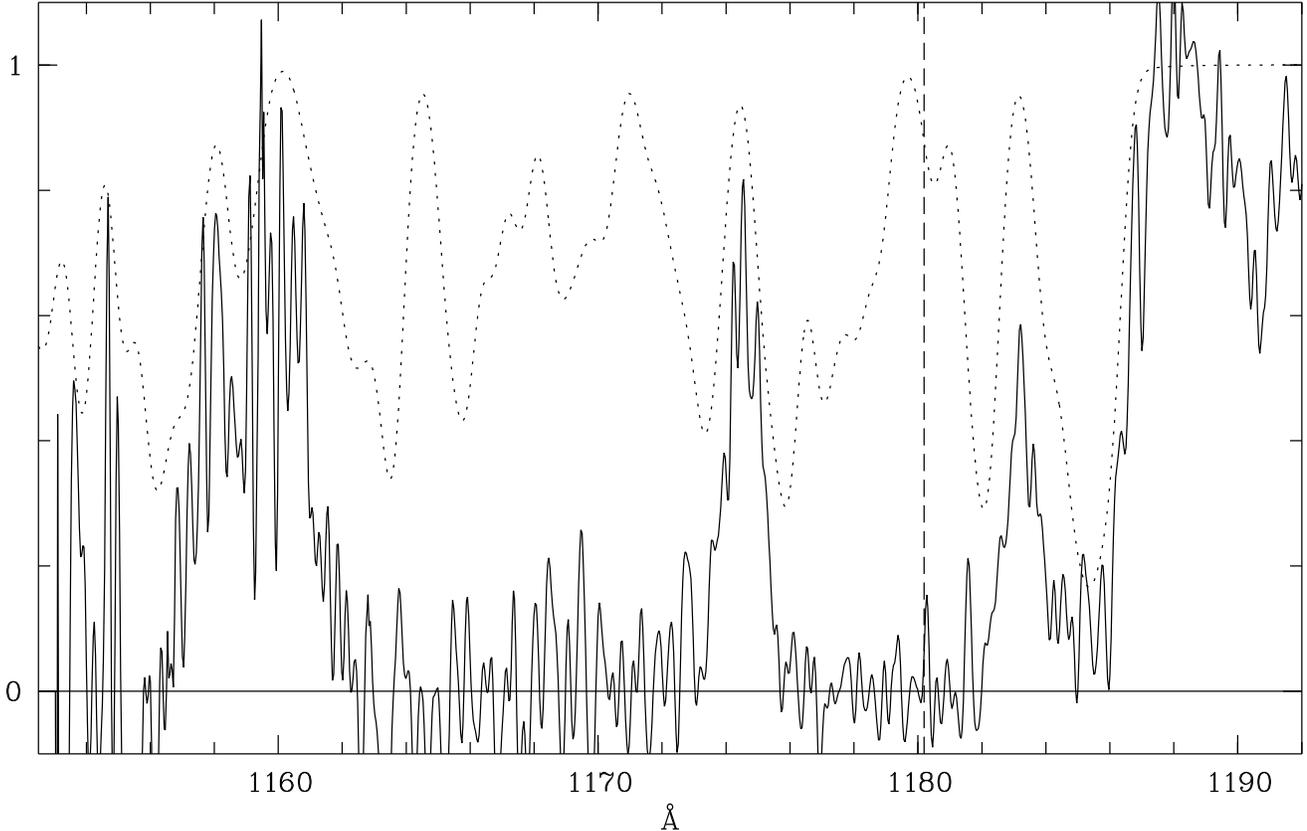}
\caption{Normalized part of the GHRS spectrum overlaid with a model 
He\,{\sc ii} absorption spectrum 
predicted on the basis of redshifts, H\,{\sc i} column densities and 
velocity dispersion parameters from  Ly$\alpha$ forest lines in optical data.
 The dotted  curve corresponds to the assumptions 
$N_{\subs{He\,{\sc ii}}}/N_{\subs{H\,{\sc i}}}=45$  and pure thermal broadening $b_{\subs{He\,{\sc ii}}}=0.5\ 
b_{\subs{H\,{\sc i}}}$. }
\label{feta45}
\end{figure*}

Assuming instead $\eta\simeq\,45$ (Haardt \& Madau 1996) and pure thermal 
broadening of
He\,{\sc ii}, for which there is evidence from theoretical Ly$\alpha$
forest simulations (e.g.\ Zhang et al.\ 1997), the known Ly$\alpha$
forest clouds in the 1160 \AA\ 
void contribute only $\sim$ 50 \%
of the observed opacity and an additional optical depth $\tau^{\subs{GP}}_{\subs{He\,{\sc
ii}}}\simeq\,0.3$ 
is required (see Fig.~\ref{feta45}). The latter includes the contribution by faint optically thin forest 
lines not detected in our high-resolution spectrum.
According to Madau \& Meiksin (1994)
the  optical depth  $\tau_{\subs{GP}}$  of the diffuse He\,{\sc ii} gas is related to
the background intensity in photoionization equilibrium by
\begin{equation}
\Omega_{\subs{diff}} = 0.25 \cdot \tau_{\subs{GP}}^{0.5} 
\cdot h^{-1.5}_{50} \cdot S^{-0.5}_{\subs{L}}
\cdot J^{0.5}_{-21}
\end{equation}
where  $J_{-21}$
is the mean intensity of the ionizing background at 912~\AA\
 in units of 10$^{-21}$ erg\,cm$^{-2}$\,s$^{-1}$\,\AA$^{-1}$ sterad$^{-1}$,
h$_{50}$ is the Hubble constant in units of H$_{0}$ = 50 km\,s$^{-1}$\,Mpc$^{-1}$ and
q$_{\subs{0}}$=0.5.
With $\tau_{\subs{GP}}^{\subs{He\,{\sc ii}}}=0.3$, $J_{-21}=0.5$, $\eta=45$
($S_{\subs{L}}=\eta$/1.8) 
Eq.\ (2) yields $\Omega_{\subs{diff}}=0.02$ h$^{-1.5}_{50}$. 
Because of the assumed lower limit to a He\,{\sc ii} forest contribution, this must 
be considered as a strict upper limit.

Our result for the void at 1160 \AA\ is also consistent with the diffuse gas density derived by 
Hogan et al.\ (1997) for the void in Q\,0302$-$003  at $z=3.17$. They found 
2 $\leq \tau \leq$ 1.3 for the diffuse component. With the same parameters 
used in this paper this translates to 0.043 h$^{-1.5}_{50} \geq 
\Omega_{\subs{diff}} \geq$ 0.035 h$^{-1.5}_{50}$ for \mbox{$\eta$ = 45} and 
0.029 h$^{-1.5}_{50} \geq \Omega_{\subs{diff}} \geq$ 0.023 h$^{-1.5}_{50}$ for 
$\eta$ = 100, respectively. The latter  softer ionizing spectrum 
appears more appropriate at $z=3.17$ (Songaila \& Cowie 1996) and is 
consistent with the diffuse density $\Omega \simeq$ 0.02 h$^{-1.5}_{50}$ 
derived at $z=2.82$ in HE\,2347$-$4342. With the same shape of the ionizing background at $z=2.8$ and $z=3.17$, the diffuse density found in the HE\,2347$-$4342 void is lower by a factor of 2 than in the Q\,0302$-$003 void.

If the voids are caused by additional ionizing sources -- and we have no better
explanation for voids in the Ly$\alpha$ forest -- the background radiation may 
be irrelevant for the ionization in the voids. In case of an AGN as ionizing
source, $\eta$ would be smaller ($\eta\leq\,10$) and the necessary amount
of diffuse gas would increase. A star dominated ionizing source, on the other 
hand would mean a larger $\eta$ ($\geq\,100$) and no diffuse medium 
would be required. We must conclude that at present we have no real constraints
for a diffuse medium in the voids.

\subsection{The ``troughs''}
The mean flux measured from our data in the wavelength range from 1175.5
 to 1181
\AA\ is $f_{\lambda}=0.023\,\pm\,0.18\ 10^{-15}$ erg s$^{-1}$ cm$^{-2}$ \AA$^{-1}$. The continuum
flux measured just longward of the He\,{\sc ii} edge is 
$f_{\lambda}=3.6\ 10^{-15}$ erg\,s$^{-1}$\,cm$^{-2}$\,\AA$^{-1}$. The corresponding flux ratio implies a high optical depth $\tau=5$.
When the flux depression   by discrete He\,{\sc ii} absorption of Ly$\alpha$ forest clouds is
taken into account the minimum required optical depth is 
$\tau\approx\,4.8_{-2}^{+\infty}$
 to explain the vanishing flux.

If the same ionizing UV background is responsible for the ionization in
the ``troughs'' the density of the diffuse gas can easily be estimated.
According to Eq.~2, $\tau\,> 4.8\ (2.8)$
 yields $\Omega_{\subs{diff}}> 0.077\ (0.059)$
h$^{-1.5}_{50}$ ($\geq\,0.03$ (0.023) h$^{-1.5}_{50}$ for $\eta= 250$)
close to the total big bang nucleosynthesis baryon
density $\Omega_{\subs{bb}}$ = 0.05 h$^{-2}_{50}$ (Walker et al.\ 1991).
Assuming that the He\,{\sc ii} opacity in the ``troughs''\- represents the
normal IGM at $z=2.85$ and the ``void'' at 1160 \AA\ is caused
by higher ionization due to additional local ionizing sources, we are forced
to conclude that the baryons are largely in the diffuse or at
least the low density part of the IGM. This would be at variance
with all recent hydrodynamical simulations of hierarchical structure
formation (Cen et al.\ 1994; Miralda-Escud\'{e} et al.\ 1996;
Zhang et al.\ 1997; Rauch et al.\ 1997;
Meiksin 1997) which predict that except a few percent all of the
baryons are in the Ly$\alpha$ forest clouds, i.e.\ fragmentation
of the IGM should be nearly complete.

There is a further argument against photoionization of the trough 
component by the same ionizing radiation as the voids. Since 
$\tau($H\,{\sc i})$=4\,\tau($He\,{\sc ii})/$\eta$, the 
Gunn-Peterson constraints $\tau($He\,{\sc ii})\,$\geq\,2.8$ (4.8) and
$\tau($H\,{\sc i})\,$\leq\,0.05$ lead to $\eta\geq\,220$ (380), which is
inconsistent with an AGN dominated background at $z\leq\,3$ for which there 
is ample evidence (e.g.\ Haardt \& Madau 1996). 

 Trying  to explain 
the observed flux depression, e.g.\ at 1180 \AA,  
we also estimated the contribution from  further absorber candidates like
 a) He\,{\sc ii} absorption from weak Ly$\alpha$ clouds which are not
detectable in our data, b) higher Lyman series lines from Ly$\alpha$ clouds at low redshifts,
c) neutral helium absorption or d) heavy element absorption lines. Of course none of these
absorption lines  alone can account for the total absorption observed.

In our calculation we considered   at the same time a) a diffuse He\,{\sc ii}
 absorbing component with $\tau$=0.2, b)  He\,{\sc i}\,584
absorption from the strong Ly$\alpha$ clouds detected at $z=1$, c) He\,{\sc ii}
absorption lines from presumed 
weak Ly$\alpha$ forest clouds with log\,$N_{\subs{H\,{\sc i}}}=12$ adopting $\eta=100$ in addition to  
 He\,{\sc ii} absorption from detected Ly$\alpha$ forest clouds. 
Unless one assumes
$N_{\subs{He\,{\sc ii}}}/N_{\subs{H\,{\sc i}}}=1000$ all these absorbers
cannot explain the vanishing flux.

Further absorption  turning on just at the He\,{\sc ii} edge is expected
  by the strong O\,{\sc iii}
doublet at 303 and 305 \AA.
Down to $z_{\subs{abs}}=2$ it is possible to identify MLS by C\,{\sc iv} absorption longward of
Ly$\alpha$  emission in the optical data. Besides the associated system we
identified 9 MLS, which  often split into several components.
For absorbers with $z>2.8$ the O\,{\sc iii}\,303,305 doublet  falls longward of 1150 \AA.
We calculated simple CLOUDY (Ferland 1993) models to estimate their influence. Unless
adopting
very unrealistic high column densities for O\,{\sc iii} these lines cannot
explain a
total absorption.

\subsection{Delayed He\,{\sc ii} ionization}
We see only one way to resolve the
observed inconsistencies: He\,{\sc ii} ionization in the ``trough'' component
has to be incomplete.
Meiksin \& Madau (1993) have predicted H\,{\sc i} absorption ``troughs'' with
vanishing flux, velocity broadened by the Hubble expansion. This has
so far not been seen in H\,{\sc i} up to $z=4.9$.
We propose here that we observe a patchy intergalactic medium in
which the reionization of the universe is still incomplete at 
 $z\simeq$ 3 for the second ionization of He. This possibility has
been predicted by various authors, e.g.\ Miralda-Escud\'{e} \&
Rees (1993) and Madau \& Meiksin (1994). The reason for the delayed
ionization of He\,{\sc ii} is the  much smaller number of He\,{\sc ii} ionizing
photons compared to H ionizing photons. Madau \& Meiksin (1994) show
that with QSO spectra varying as $\nu ^{-1.9}$ and the quasars turned
on at $z=6$, hydrogen is fully ionized at $z=5$, but that He\,{\sc ii} is only
partially ionized at $z=3$. According to recent work by 
Shaver et al.\ (1996), QSOs are an order of magnitude less abundant
at $z>4$ than at $z <3$, i.e. the turnon of the QSOs as ionizing
sources is even later than assumed by Madau \& Meiksin (1994).
Furthermore, there is evidence from modelling of the Ly$\alpha$
forest that for $z>3$ the UV-background due to QSOs is not
sufficient and that additional sources of photons are required for $z>3$
(Rauch et al.\ 1997). If these sources are stars, the resulting softer UV
background possibly meets the condition for delayed He\,{\sc ii} ionization.
In addition, it has been shown by Giroux \& Shapiro (1996)
that in case of an IGM partially collisionally ionized due
to bulk heating in addition to photoionization, He ionization will
be lag behind hydrogen ionization.
There is also evidence of an  abrupt change in the
C\,{\sc iv}/Si\,{\sc iv} column density ratio  around
$z=3$ (Songaila \& Cowie 1996), which indicates a softening of the
ionizing UV background with increasing $z$:  
$\eta\approx\, 20$ to 40 below $z=3.1$ and $\eta\,> 250$ above $z=3.1$.

In case of only singly ionized helium, the amount of diffuse gas
needed to produce the ``trough'' opacity $\tau_{\subs{GP}}^{\subs{He\,{\sc ii}}} \geq\,4.8$
is given by
\begin{equation}
\tau _{\subs{GP}}^{\subs{He\,{\sc ii}}} (z=2.8) = 3.6 \cdot 10^{4} 
\cdot \Omega_{\subs{diff}} \cdot
h_{50} 
\end{equation}
for q$_{\subs{0}}=0.5$ (Madau \& Meiksin 1994)
which leads to $\Omega_{\subs{diff}} \geq$ 1.3 $\cdot$ 10$^{-4}$ h$^{-1}_{50}$.
Only 0.3 \% of the big-bang nucleosynthesis baryon density $\Omega_{\subs{bb}}$
= 0.05 h$^{-2}_{50}$ is required to produce the blacked-out troughs.
This would be consistent with the theoretical predictions that
fragmentation of the IGM is nearly complete.

It is easy to show that even for a soft ionizing spectrum ($\alpha$ = 2)
the ``troughs'' will not be seen in He\,{\sc i}\,584 \AA\ since
$N_{\subs{He\,{\sc i}}}/N_{\subs{He\,{\sc ii}}}
\simeq\, 2\, \cdot\, 10^{-7}$ for $\Omega$ = 0.02 (cf.\ Jakobsen 1995). The FOS spectrum
around 2240 \AA\ confirms the absence of a He\,{\sc i} trough.

\subsection{No proximity effect?}
An extremely luminous QSO like HE\,2347$-$4342 is expected to show a proximity effect
except for the rather improbable case that the QSO has been turned on only
recently.

The influence of the UV radiation of the QSO itself on the second
ionization of He has been studied theoretically by various authors
(e.g.\ Zheng \& Davidsen 1995; Meiksin 1995; Giroux et al. 1995).
The prediction is that close to the QSO He\,{\sc ii} is additionally ionized
and that the effect is stronger in diffuse than in clumped gas.
Consequently, if the He\,{\sc ii} opacity close to the QSO is dominated by
a diffuse component, we expect increased transparency in a He\,{\sc iii} bubble
around the QSO, close to the rest redshift of the QSO, while the
effect will be weaker in the Ly$\alpha$ forest.
No proximity effect has been seen in HS\,1700+6416 (Davidsen et al.\ 1996), 
while Hogan et al.\ (1997)  observed in Q\,0302$-$003 a ``shelf''
in the He\,{\sc ii} opacity \mbox{$\sim$ 12 \AA} blueward of the main He\,{\sc ii} edge which
they interpret as a proximity effect in diffuse He\,{\sc ii} gas.
If, as concluded above, the ``troughs'' in the He\,{\sc ii} forest at 1162--1173 
\AA\ and 1176--1182 \AA\  are caused by not fully ionized gas, we should
expect that HE\,2347$-$4342, which  at the time of observation is one
of the most luminous sources in the universe,
 produces a strong proximity effect.

In the following we  estimate the He\,{\sc ii} ionizing flux of
HE\,2347$-$4342 and discuss its effect on the surrounding medium.
UV spectra of high-redshift quasars are strongly influenced by absorption of
intervening absorbers. In order to find the intrinsic spectral energy
distribution of the quasar corrections have to be applied to the observed
data.  In the following we will consider only the flux depression by 
cumulative hydrogen continuum absorption of intervening absorbers, 
neglecting effects of dust absorption along the line of sight or line
blanketing in the UV. Uncertainties in the  absolute flux calibrations 
of the spectra up to 10--20 \% are possible.

The dereddened spectra of HE\,2347$-$4342 were corrected for continuum absorption by
neutral hydrogen in the strong LLS at $z=2.739$. The continuously rising spectrum
of HE\,2347$-$4342 
shows no evidence for further strong absorbers with log\,$N_{\subs{H\,{\sc i}}}>16$.
 The additional ``Lyman valley'' depression of the quasar continuum due to the
cumulative hydrogen absorption by the numerous Ly$\alpha$ clouds with
$12\leq\,$log\,$N_{\subs{H\,{\sc i}}}\leq\,16$ was estimated by Monte Carlo simulations.
 Since
$\lambda_{\subs{rest}}=228$ \AA\ is not directly observable,  we take the corrected continuum value at the
smallest observed wavelength ($\lambda_{\subs{rest}}=298$ \AA), assuming that it is close to the
flux at 228 \AA. We find 
$f_{\lambda}$(885.78 \AA)\,$\approx\,f_{\lambda}$(1157 \AA)\,=
6.4\ 10$^{-15}$ erg s$^{-1}$ cm$^{-2}$ \AA$^{-1}$, i.e. nearly a factor of 2 higher
than the observed flux. 

For the QSO flux at $\lambda_{\subs{rest}}=912$ \AA\ we considered the  flux
in the range from 3560 to 3580 \AA\ in the optical data just longward of the corresponding
but unobserved wavelength $3.885\times\,912=3543$ \AA.
Since in low-resolution optical data the flux is depressed by unresolved line
blanketing of Ly$\alpha$ forest lines, we rebinned our high-resolution optical   
 data  to estimate the flux depression by line blending. The resulting flux
is then $f_{\lambda}(3543$ \AA)=1.75 10$^{-15}$ erg s$^{-1}$ cm$^{-2}$ \AA$^{-1}$
yielding a ratio of $S_{\subs{L}}=f_{\nu}(912)/f_{\nu}(228)=4.4$.
 Adopting a pure power law ($f_{\nu}\propto\,\nu^{\alpha}$)   for the intrinsic
 EUV spectral energy distribution 
 $S_{\subs{L}}=4.4$ turns into $\alpha=-1.0$, which is a typical value for several 
luminous QSOs observed with HST (K\"ohler \& Reimers 1996).
In the proximity of the quasar we thereby do not expect a strong contribution from diffuse Helium.
With 
\begin{equation}
\tau_{\subs{GP}}^{\subs{HeII}}\sim\,0.45\,S_{\subs{L}}\,\tau_{\subs{GP}}^{\subs{HI}}
\end{equation}
 (Madau \& Meiksin 1994)
and $\tau_{\subs{GP}}^{\subs{HI}}<0.05$ (Giallongo et al.\ 1994) 
we expect for $S_{\subs{L}}=4.4$
an optical depth of the diffuse component 
$\tau_{\subs{GP}}^{\subs{HeII}}=0.1$ comparable to $\tau_{\subs{GP}}^{\subs{HeII}}\leq\,0.3$ derived from the observations. 
For the associated system we would expect $N_{\subs{He\,{\sc ii}}}/N_{\subs{H\,{\sc i}}}=8$
($\eta=1.8\,S_{\subs{L}}$) also consistent with the
observation  at least for the  components at highest redshifts when adopting
turbulent line broadening.

For a QSO flux at the He\,{\sc ii} edge of $f_{\nu}=1.7\ 10^{-27}$ erg
s$^{-1}$ cm$^{-2}$ Hz$^{-1}$ at $z=2.885$ (for $f_{\lambda}(1157$ \AA)=6.4 10$^{-15}$ erg
s$^{-1}$ cm$^{-2}$ \AA$^{-1}$ as derived above and Eq.~6 from Giroux
et al.\ 1995) we get $f_{\subs{q}}=1.1\ 
10^{-21}$ erg  s$^{-1}$ cm$^{-2}$ Hz$^{-1}$ at $z_{\subs{abs}}=2.819$ (corresponding 
to 
He\,{\sc ii}\,304 absorption at 1160 \AA) which has
to be compared with a background flux
$f_{\nu}=J_{\nu}\cdot\,4\pi=2.5\ 10^{-22}$ erg s$^{-1}$ cm$^{-2}$ Hz$^{-1}$
(Haardt \& Madau 1996) or $f_{\nu}=1.5\ 10^{-21}$ erg s$^{-1}$ cm$^{-2}$
Hz$^{-1}$ (Bechtold 1995 with $S_{\subs{L}}=f_{912}/f_{228}=25$).
This means that HE\,2347$-$4342 with its observed hard ($\alpha=-1$) EUV
spectrum dominates He\,{\sc ii} ionization at least up to the $\lambda\,1160$
\AA\ void. This is not observed.
 Why does HE\,2347$-$4342
ionize He\,{\sc ii} in the associated system but not in the troughs, in
particular the region $\lambda\lambda1176$--1180 \AA? There is only one
plausible reason for the nonexistence of a proximity effect in He\,{\sc ii}:
shielding by clouds optically thick in the He\,{\sc ii}\,228 \AA\ continuum 
along 
the line of sight. Possible candidates are the associated system clouds.

A reliable estimate of the neutral hydrogen column densities of the individual
components of the associated system is difficult due to 
the complex structure of the system and saturation effects. Higher Lyman series lines up to
H\,{\sc i}\,920 are visible, but the signal to noise ratio of our spectrum 
decreases rapidly
at shorter wavelengths. According to absorption line fits the 
largest column density for a single component 
should not exceed log\,$N_{\subs{H\,{\sc i}}}\approx\,16$  for $b_{\subs{min}}=20$ km s$^{-1}$.
From  optical data we cannot preclude the existence of a small break at the
Lyman edge expected at $\approx$ 3550 \AA. The maximum total H\,{\sc i} column density compatible
with the optical data is then $\approx\,6\ 10^{16}$ cm$^{-2}$ . With the H\,{\sc i} column densities derived
from  modelling of Lyman series lines adopting $b>20$ km s$^{-1}$  for the 14 individual components the observed He\,{\sc ii} absorption
requires  $\eta>45$ for thermal line broadening. Thus for a
 single component with log\,$N_{\subs{H\,{\sc i}}}=16$ and $\eta>45$ we find 
$N_{\subs{He\,{\sc ii}}}>4.5\ 10^{17}$ cm$^{-2}$. The clouds become optically
thick for He\,{\sc ii} at a column density 6.3 10$^{17}$ cm$^{-2}$ and self
shielding could enhance the He\,{\sc ii} absorption.
In fact, as was mentioned in Sect.~3, $\eta=100$ gives a better fit for 
the lower redshift components ($z=2.8896$--2.8985) which leads to possibly $N_{\subs{He\,{\sc ii}}}\geq\,10^{18}$ cm$^{-2}$, sufficient to shield the IGM in our direction from He\,{\sc ii} ionizing photons.
 A detailed
analysis of the  associated system
with better optical data will help to improve the constraints for 
 the He\,{\sc ii} continuum
absorption.

What is the origin of the 1160 \AA\ and 1174 \AA\ voids? If the troughs are
due to unionized He\,{\sc ii} which completely shields the EUV background at
$\lambda\leq\,228$ \AA, we have to assume that individual sources (QSO and/or
starburst galaxies) create He\,{\sc iii} regions within the unionized
IGM. With a diameter of 2.8 h$_{50}^{-1}$ Mpc for the 1160 \AA\ void, the
requirements on the luminosities of  possible sources are moderate (cf.\ Bajtlik et al.\ 1988,
Giroux \& Shapiro 1996). The observed $\eta\geq\,45$ would favour at least a
contribution by stars in the ionizing source.

\section{Summary and conclusions}

We have shown that the IGM in the line of sight of HE\,2347$-$4342 is
``patchy'' in the sense that there are \mbox{``voids''} and ``troughs''.
In the voids, the He\,{\sc ii} opacity can be understood as Ly$\alpha$
forest line opacity with either $\eta$ = 100 and  
$b_{\subs{He\,{\sc ii}}} = b_{\subs{H\,{\sc i}}}$ 
and no diffuse medium  or Ly$\alpha$ forest line opacity
with $\eta$ = 45 and $b_{\subs{He\,{\sc ii}}} = 0.5\ b_{\subs{H\,{\sc i}}}$
 plus a diffuse medium
with $\tau_{\subs{GP}}$ = 0.3.
The He\,{\sc ii} opacity seen in the voids is consistent with what has been
found in HS\,1700+6416 (Davidsen et al.\ 1996) or in the Q\,0302$-$003
void by Hogan et al. (1997).
In the ``troughs'' we need in addition to the He\,{\sc ii} Ly$\alpha$ forest
opacity a continuous opacity of $\tau_{\subs{GP}}^{\subs{He\,{\sc ii}}}$ $>$ 4.8 ($\tau\,> 2.8$ is a strict lower limit). 
A natural explanation for this dichotomy  is the
assumption of delayed He\,{\sc ii} ionization, as predicted by
e.g.\ Madau \& Meiksin (1994), where in the phase of reionization of the
universe there are still not yet ionized He\,{\sc ii} regions between the expanding
He\,{\sc iii} regions. In that case $\Omega_{\subs{diff}}\geq\,1.3\,\cdot\,10^{-4}$ 
h$^{-1}_{50}$ would be necessary.
The absorption troughs have the shape and widths as predicted by
Meiksin \& Madau (1993).

The patchy He\,{\sc ii} opacity appears to be the ``missing link''
between the $\tau=1$ at $\bar{z}=2.4$ in HS\,1700+6416 and the high
opacity in Q\,0302$-$003 at $z=3.2$: The ``voids''\\ in HE\,2347$-$4342
have optical depths comparable to the mean opacity observed in HS\,1700+6416
and represent the reionized IGM, while the ``troughs'' represent the IGM
with delayed He\,{\sc ii} reionization which may also be the case for 
Q\,0302$-$003 and PKS\,1935$-$692 at $z\ga\,3.1$. It is important to
repeat the He\,{\sc ii} opacity measurement in case of Q\,0302$-$003 
with STIS because the background subtraction procedure applied to the
GHRS data appears doubtful at such low flux levels.

Our He\,{\sc ii} observations are in accordance with Songaila and Cowie's (1996)
finding from the observed C\,{\sc iv}/Si\,{\sc iv} ratio as a function of
 redshift 
that around $z = 3$ there appears to be a transition from a soft ionizing
UV background ($z > 3$) to relatively hard ionizing radiation ($z < 3$).
If confirmed by further observations, the delayed He\,{\sc ii} ionization gives
important constraints on the evolution of the UV background for $z > 3$. 

The same ionization model of Madau \& Meiksin (1994) 
which predicts incomplete He\,{\sc ii} ionization for $z\geq\,3$ 
predicts incomplete H\,{\sc i} ionization for $z\geq\,4.9$. After our 
observation of delayed He\,{\sc ii} ionization there appears to be a 
realistic chance to observe directly the phase of reionization of 
the universe if $z\geq\,5$ QSOs exist since the latter might show H\,{\sc i} 
blacked-out troughs similar to what we have seen in He\,{\sc ii}.
One has to keep in mind, however, that HE\,2347$-$4342 is just one line of
sight, and that before generalizations are possible, more objects have to
be probed.

The absence of a measurable proximity effect in HE\,2347$-$4342
 might be explained by shielding of $\lambda< 228$~\AA\ radiation by the 
clouds responsible for the strong associated system seen toward this quasar.

Observations of the He\,{\sc ii} troughs in HE\,2347$-$4342 with STIS
in low resolution can be expected to yield an improved lower limit to the 
diffuse density in the yet unionized medium, and HE\,2347$-$4342 is
even bright enough in the 1160 \AA\ void for FUSE to resolve directly the 
He\,{\sc ii}\,304~\AA\ forest.

\begin{acknowledgements}
This work has been supported by the Verbundforschung of BMBF under No.\ 50\,OR\,96\,016. We also like to thank Hans-Emil Schuster, Bo Reipurth and Guido and Oscar Pizarro for taking Schmidt plates for the QSO survey over many years.
We are indebted to Peter Jakobsen who critically read the manuscript and 
helped to improve it.

\end{acknowledgements}
{} 

\end{document}